\makeatletter
\@namedef{ver@subfig.sty}{9999/99/99}
\@namedef{opt@subfig.sty}{labelformat=empty}
\makeatother

\documentclass[twocolumn,APA]{USG} 

\usepackage{anyfontsize} 
\usepackage{colortbl}   
\usepackage{xcolor}     
\usepackage{steinmetz}
\usepackage{tabularx}
\usepackage{subcaption}
\usepackage[authoryear]{natbib}

\graphicspath{{./images/}}

\articletype{ORIGINAL ARTICLE}%

\received{--}
\revised{--}
\accepted{--}
\journal{arXiv}
\volume{0}
\startpage{1}
\articledoi{}

\begin{document}

\title{Evaluating multimodal emotion recognition in proactive conversational agents: A user study}

\transtitle{}
\subtranstitle{}

\author[1]{Adnana Dragut}
\author[1,2]{Raquel Lacuesta}[https://orcid.org/0000-0002-4773-4904]
\author[3]{F. Xavier Gaya-Morey}[https://orcid.org/0000-0003-1231-7235]
\author[3]{Jose M. Buades-Rubio}[https://orcid.org/0000-0002-6137-9558]


\address[1]{\orgdiv{Escuela Universitaria Politécnica de Teruel}, \orgname{Universidad de Zaragoza}, %
\orgaddress{C/ Atarazana, 2, \city{Teruel}, \postcode{44003}, \state{Aragón}, \country{Spain}}}

\address[2]{\orgdiv{I3A (Institute of Engineering Research of Aragon)}, \orgname{Universidad de Zaragoza}, %
\orgaddress{C. de Mariano Esquillor Gómez, s/n, \city{Zaragoza}, \postcode{50018}, \state{Aragón}, \country{Spain}}}

\address[3]{\orgdiv{Universitat de les Illes Balears}, \orgname{Universitat de les Illes Balears}, %
\orgaddress{Carretera de Valldemossa, km 7.5, \city{Palma}, \postcode{07122}, \state{Illes Balears}, \country{Spain}}}

\corres{F. Xavier Gaya-Morey (\email{francesc-xavier.gaya@uib.es})}



\transkeywords{}

\abstract[ABSTRACT]{This article presents a multimodal emotion recognition module integrated into a proactive Socially Interactive Agent (SIA) powered by generative artificial intelligence. The system evaluates real-time affective states through two distinct channels: a computer vision-based facial recognition module and a semantic linguistic analysis engine. To validate the framework, an empirical study was conducted with 20 users who engaged in dynamic, unscripted dialogues with the conversational agent. The findings reveal a significant discrepancy between automated visual cues and actual internal emotional states. When interacting with the AI, users consistently exhibited a ``poker face'' effect, displaying serious, concentrated facial expressions even when experiencing positive emotions. Consequently, the generative AI linguistic analysis proved significantly more reliable, by contextualizing the users' verbal expressions. Furthermore, an analysis of the interaction dynamics demonstrated that SIAs can effectively elicit specific emotions by adapting conversational themes and employing structured linguistic patterns, such as empathetic or humorous language. However, the study also noted that instances of uncalibrated proactivity occasionally led to user disengagement and a perception of artificiality. Ultimately, this research highlights the necessity of refining SIAs to dynamically adapt to users' emotional evolution, relying on deep linguistic context to foster more natural, human-like interactions.}

\transabstract[transABSTRACT]{}




\maketitle


\section{Introduction}

    Understanding and responding to emotions is a fundamental component of human communication. For Socially Interactive Agents (SIAs) to establish meaningful connections with users, they must be capable of recognizing and adapting to the user's emotional state in real time \citep{feng2022emowoz}. Integrating this affective awareness is essential for humanizing conversational agents, fostering natural dialogues, and ultimately improving their overall user acceptance \citep{gao2022emotion}.
    
    Currently, emotion detection systems rely on a variety of modalities, ranging from facial expression recognition and voice processing to gaze tracking and the analysis of biological signals, such as electroencephalograms (EEG) and electrocardiograms (ECG) \citep{siam2022deploying, garcia-magarino2018agent-based}. Recent research has demonstrated significant progress in multimodal emotion recognition applied to social agents. For example, \citet{makiuchi2021multimodal} proposed a novel cross-representation model combining speech and text that effectively outperforms traditional speech-only or text-only unimodal approaches \citep{christ2023muse}. Similarly, \citet{katada2023effects} evaluated the fusion of physiological signals with other modalities to estimate different types of sentiment during naturalistic human-agent interactions.
    
    Despite these technical advancements, most existing studies rely on pre-recorded datasets, theoretical simulations, or passive observation. There is a notable lack of research integrating multimodal emotion detection within dynamic, real-time conversations driven by generative AI. Specifically, current literature does not sufficiently address how proactive, AI-generated dialogues attempt to connect with specific user emotions, nor how these interactions directly influence the user's facial expressions and internal affective states during a live exchange.
    
    To bridge this gap, this study presents the design and empirical evaluation of a proactive conversational SIA. The proposed system integrates two core emotion detection techniques: a computer vision module for facial expression recognition and a generative AI-based linguistic analysis module. Through practical, real-time evaluations with users interacting with the SIA, this research investigates how human emotions actually manifest during unscripted interactions. By comparing automated visual data, semantic linguistic analysis, and subjective user self-reports, this study highlights critical interaction patterns and evaluates the real-world accuracy of these emotion detection methods.
    
    The remainder of this article is organized as follows: Section \ref{sec:related-work} reviews related work in emotion recognition and SIAs. Section \ref{sec:RQ} outlines the research objectives and specific research questions. Section \ref{sec:methodology} details the methodology, including the system design and evaluation procedure. Section \ref{sec:results} presents the objective results of the multimodal evaluation. Section \ref{sec:discussion} discusses these findings and answers the core research questions. Section \ref{sec:limitations} acknowledges the study's limitations and outlines future work, and Section \ref{sec:conclusion} provides the final conclusions.

\section{Related Work}
\label{sec:related-work}

    The success of interactions between humans and artificial agents relies heavily on emotional identity and the understanding of cultural affect. These elements allow agents to maintain emotional coherence, transforming their perception from mere machines into entities capable of establishing meaningful connections \citep{hoey2016affect, malhotra2021emotions}. As argued by \citet{joby2022effect}, the ability of agents to engage in emotional contagion, the transmission of affective states between participants, is vital for enhancing trust, empathy, and prosocial orientation during social exchanges.
    
    \subsection{Emotional Intelligence in Socially Interactive Agents}
    
        SIAs are increasingly integrated into diverse domains, necessitating the incorporation of emotional intelligence to improve human-machine dynamics. Emotions are not merely aesthetic; they are essential for creating realistic behaviors in social simulations by integrating cognition and social relationships. For instance, \citet{alanazi2023prediction} focused on simulating emotions to foster empathy, thereby enhancing agent autonomy and adaptability. Similarly, \citet{samsonovich2014developing} explored emotionally intelligent virtual agents through biologically inspired cognitive architectures to replicate human social behavior.
        
        Further research by \citet{erol2020artificial} emphasizes the importance of recognizing human emotional states to improve bonding, proposing perception architectures for human-robot interaction. In parallel, \citet{tavabi2019multimodal} investigated multimodal deep neural networks to identify opportunities for empathetic responses. Despite these advances, some authors argue that a significant gap remains: most existing models focus on theoretical simulations and Affect Control Theory without addressing the challenges of real-time processing in dynamic interactions \citep{malhotra2021emotions, hoey2016affect}. Many studies, such as those by \citet{cipresso2012real} and \citet{hortensius2018perception}, primarily review how humans perceive emotions in agents but do not delve into how these perceptions can be processed and responded to by the agents themselves in live settings.
        
    \subsection{Facial Expression Recognition and its Limitations}
        
        Facial Expression Recognition (FER) is a fundamental pillar for detecting user states. Early deep learning approaches, such as those discussed by \citet{kalyani2023smart}, highlighted the potential of analyzing expressions alongside speech and text. \citet{wang2020human} demonstrated that bimodal approaches, combining facial features with speech, consistently outperform unimodal systems. Modern FER systems use sophisticated models like cGANs \citep{deng2019cgan}, semantic-rich frameworks \citep{chen2022semantic-rich}, and efficient architectures like SwishNet \citep{dar2022efficient-swishnet} to facilitate natural interaction.
        
        However, the transition from laboratory settings to ``in-the-wild'' applications presents severe challenges. \citet{rani2014emotion} and \citet{dalvi2021survey} point out that variations in lighting, facial angles, and cultural diversity require more robust models. While recent developments like LiteFer \citep{yang2024litefer} and video representation learning \citep{strizhkova2024video} aim to improve efficiency on limited-resource devices, a deeper technical unreliability persists. \citet{cabitza2022unbearable} and \citet{kusal2024understanding} have criticized the inconsistency of automated FER when dealing with diverse observer samples. Crucially, \citet{wang2024survey} and \citet{samadiani2019review} suggest that serious or neutral facial expressions, common in social interactions, often lead to unreliable detection. This ``ambiguity of expression'' suggests that relying solely on visual data may be insufficient for SIAs in real-world social contexts.
        
    \subsection{Text-Based Emotion Detection and Generative AI}
        
        Text-Based Emotion Detection (TBED) has become a critical component for developing empathetic agents \citep{kusal2024understanding, kusal2022review}. While it has proven valuable for big data analytics in social media \citep{kusal2021ai}, applying TBED to real-time conversation introduces complexities such as handling short texts, synonyms, and reversed word order \citep{maruf2024challenges, wen2024personality-affected}. Traditional machine learning approaches have made significant progress \citep{machova2023detection}, yet full automation remains a challenge \citep{maruf2024challenges}.
        
        The emergence of Generative AI offers a paradigm shift. \citet{bertero2016real-time} proposed real-time sentiment recognition to enable dialogue systems to respond appropriately. More recently, the work of Park et al. \citet{park2023generative} on ``Generative Agents'' demonstrates that large language models (LLMs) can simulate believable human behavior and social reflections. However, as noted by \citet{ghaffarzadegan2024generative}, these feedback-rich computational models are generally used to generate behavior rather than to analyze the user's internal state during the interaction. There is a clear need for studies that utilize generative AI to interpret the fluctuations of user emotions across different topics in real-time, especially when visual cues are absent or misleading \citep{peng2020human-machine}.
        
    \subsection{Multimodal Integration and User Experience}
        
        The consensus in the field is that multimodal systems, integrating voice, face, and text, yield the highest accuracy \citep{alonso-martin2013multimodal}. \citet{katada2024collecting} even explored the use of frontal brain signals to capture ``unexpressed sentiments,'' highlighting that users often feel more than they show. \citet{ge2024modeling} further emphasized that understanding the dependency between the speaker and the sentiment is vital for context-aware recognition.
        
        The interaction itself is a reciprocal process. \citet{woo2023reciprocal} argue that SIAs must adapt their behaviors dynamically, acting as both speakers and listeners. While agents can induce emotions like happiness or anger \citep{tanioka2025dialogue, gupta2024facial, alonso-martin2013multimodal}, the complexity of emotional signals remains a barrier to a truly positive user experience \citep{samsonovich2014developing, skillicorn2019measuring}. \citet{wen2024personality-affected} suggest that incorporating personality traits can lead to more engaging responses, yet many of these findings are based on theoretical testing \citep{li2024sia-net} or lack involvement with real-time AI-generated conversational engines \citep{orlov2024real-time}.
        
        This research addresses these limitations by developing a proactive SIA that integrates an AI system to both generate and identify emotions in real-time. By comparing ``what the user says they feel'' (self-report) with visual (SilNet) and linguistic (Generative AI) analysis, we provide a critical evaluation of the ``poker face effect.'' Unlike previous research focused on simulations, our work emphasizes a practical application that explores how linguistic analysis can capture the nuances of human emotion that facial recognition systems might miss.

\section{Research Objectives and Questions}
\label{sec:RQ}

    To address the challenges of emotional alignment in human-agent interaction, this study evaluates a multimodal framework designed to capture user affect during proactive social dialogues. Unlike previous research focused on passive recognition, we investigate the interaction dynamics when the agent takes the initiative, powered by generative AI. Our primary goal is to analyze the discrepancy between AI-recognized emotions and the users' subjective perceptions in an unscripted, dynamic environment.

    Specifically, this study seeks to answer the following two core research questions:
    
    \begin{itemize}
        \item \textbf{RQ1 (User Affective Experience): What is the emotional experience of users when interacting with a proactive socially interactive agent?} \\
        This question seeks to uncover the genuine internal impressions of the users during unscripted interactions. By relying on user self-reports and questionnaires, the objective is to establish an objective ``ground truth'' of how users genuinely feel when participating in AI-driven social dialogues.
        
        \item \textbf{RQ2 (Emotion Detection Efficacy): How effectively can multimodal systems recognize users' emotions during these interactions, and how do visual and linguistic modalities compare?} \\
        This question aims to evaluate the practical reliability of automated emotion recognition in a live setting. It focuses on observing the correspondence between the user's physical facial expressions (visual modality), the semantic text-based emotions detected by the generative AI (linguistic modality), and the user's self-reported ground truth. The goal is to determine if one modality is more effective than the other and to identify the specific causes behind detection discrepancies.
    \end{itemize}
    
    By addressing these questions, we aim to provide a clearer understanding of the ``expressive gap'' in human-agent interaction and offer design recommendations for more socially-aware conversational systems.

\section{Methodology}
\label{sec:methodology}

    This section describes the participants in the evaluation conducted, the systems designed, and the evaluation sessions carried out to gather the results.
    
    \subsection{Participants}
    
        A total of 20 participants were recruited for this study, consisting of 13 females and 7 males. The participants' ages ranged from 22 to 90 years (M = 54.90, SD = 21.31). This sample size was deemed appropriate for an exploratory Human-Computer Interaction (HCI) study aimed at identifying interaction patterns and validating multimodal frameworks. The broad age range was intentionally selected to evaluate the system's performance.
        
        To ensure data integrity, participants were screened based on the following inclusion criteria:
        
        \begin{itemize}
        
            \item \textbf{Cognitive Function:} No known history of cognitive impairments or neurological disorders.
            
            \item \textbf{Communication:} Normal or mild speech capabilities, ensuring they could interact with the system's voice interface.
            
            \item \textbf{Sensory Abilities:} Normal or corrected-to-normal vision and hearing, allowing for an unhindered perception of the interface's multimodal feedback.
            
        \end{itemize}
        
        All participants were briefed on the study's objectives and provided informed consent prior to the sessions. Participation was entirely voluntary, and no financial compensation was provided.

    \subsection{Apparatus and Materials}
    
\begin{figure}
            \centering
            \includegraphics[width=0.4\linewidth]{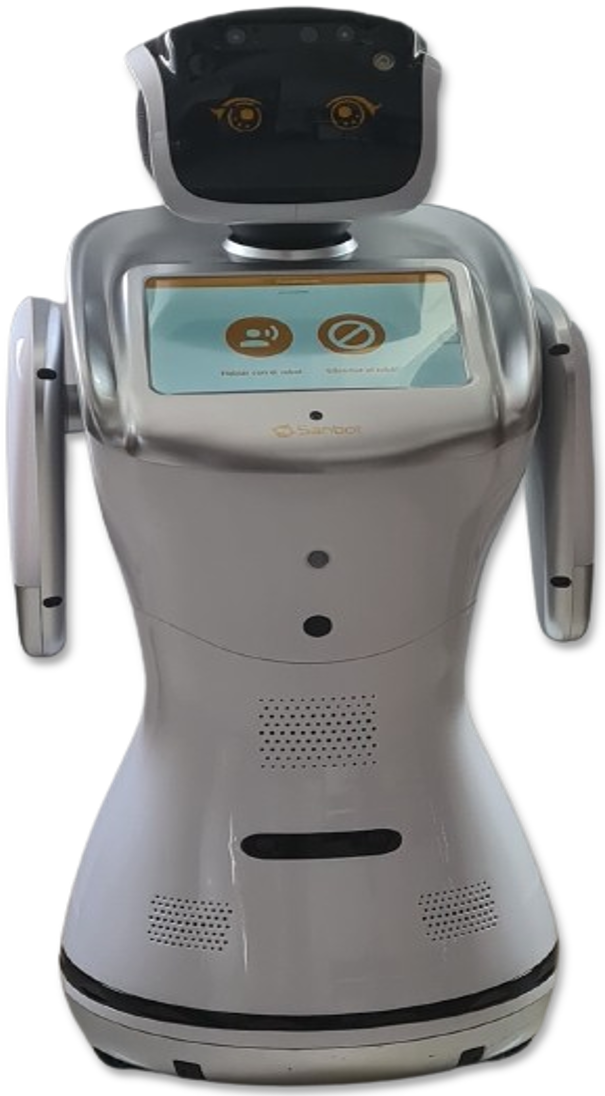}
            \caption{Sanbot Elf running the Android app.}
            \label{fig:sanbot}
        \end{figure}

        The study used the Sanbot Elf, a humanoid socially interactive agent, as the primary hardware platform (see Figure \ref{fig:sanbot}). A custom-built Android application, displayed in Figure \ref{fig:system}, was developed to serve as the middleware between the user and the agent. This application provided a minimalist visual interface to manage the session and facilitated real-time data synchronization for recording emotional states.

        The interaction took place in a controlled laboratory setting to minimize external visual and auditory noise that could interfere with the facial recognition and speech processing modules.

    \subsection{Multimodal Emotion Detection System}
    
\begin{figure*}
            \centering
            \includegraphics[width=1\linewidth]{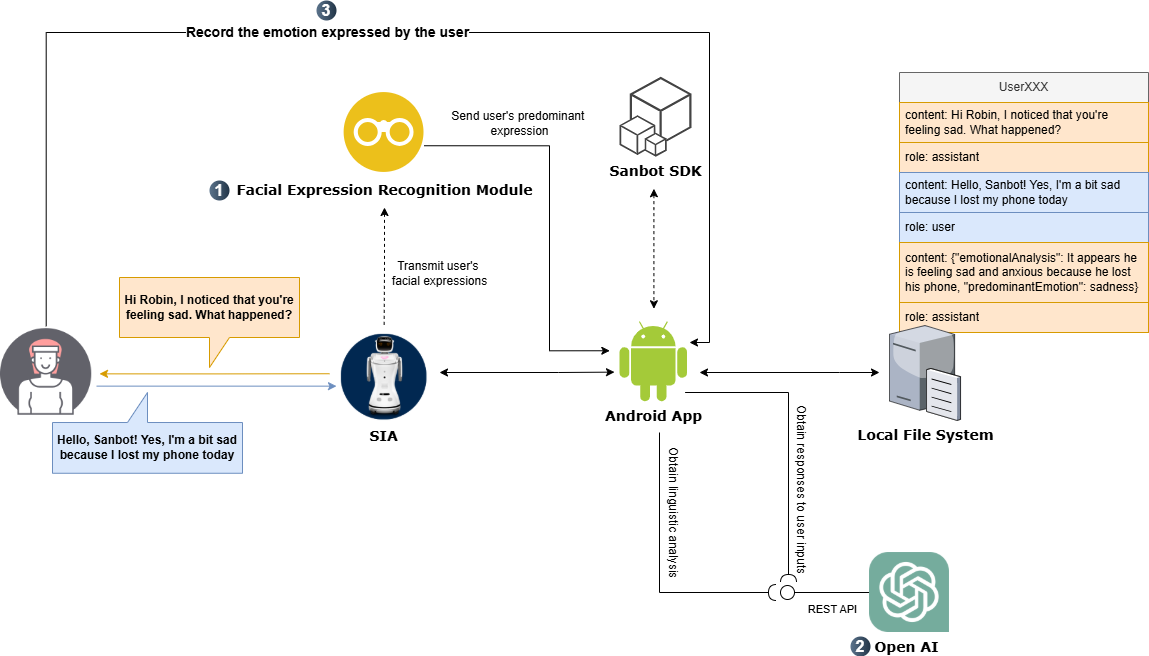}
            \caption{Component diagram of the system.}
            \label{fig:system}
        \end{figure*}

        The core of the system is a proactive AI engine designed for fluid dialogue. As shown in the component diagram (Figure \ref{fig:system}), the architecture integrates two primary emotion detection channels:

        \begin{enumerate}
        
            \item \textbf{Computer Vision Module:} This module leverages a standardized computer vision framework previously developed for SIAs \citep{gayamorey2024ai-powered}. From the various models supported by this framework, SilNet \citep{ramis2022novel} was selected due to its architectural simplicity and its specific optimization for FER tasks.
            
            \item \textbf{Linguistic Analysis Module:} The system captures the user's speech, which is then processed via the OpenAI API (GPT-4/ChatGPT). To ensure a nuanced understanding of the context, the system is programmed to perform a sentiment analysis every three interaction turns. This allows the AI to evaluate the emotional trajectory of the discourse rather than isolated words.
            
        \end{enumerate}

    \subsection{Conversational Design and Prompting}
    
\begin{figure*}
            \centering
            \includegraphics[width=\linewidth]{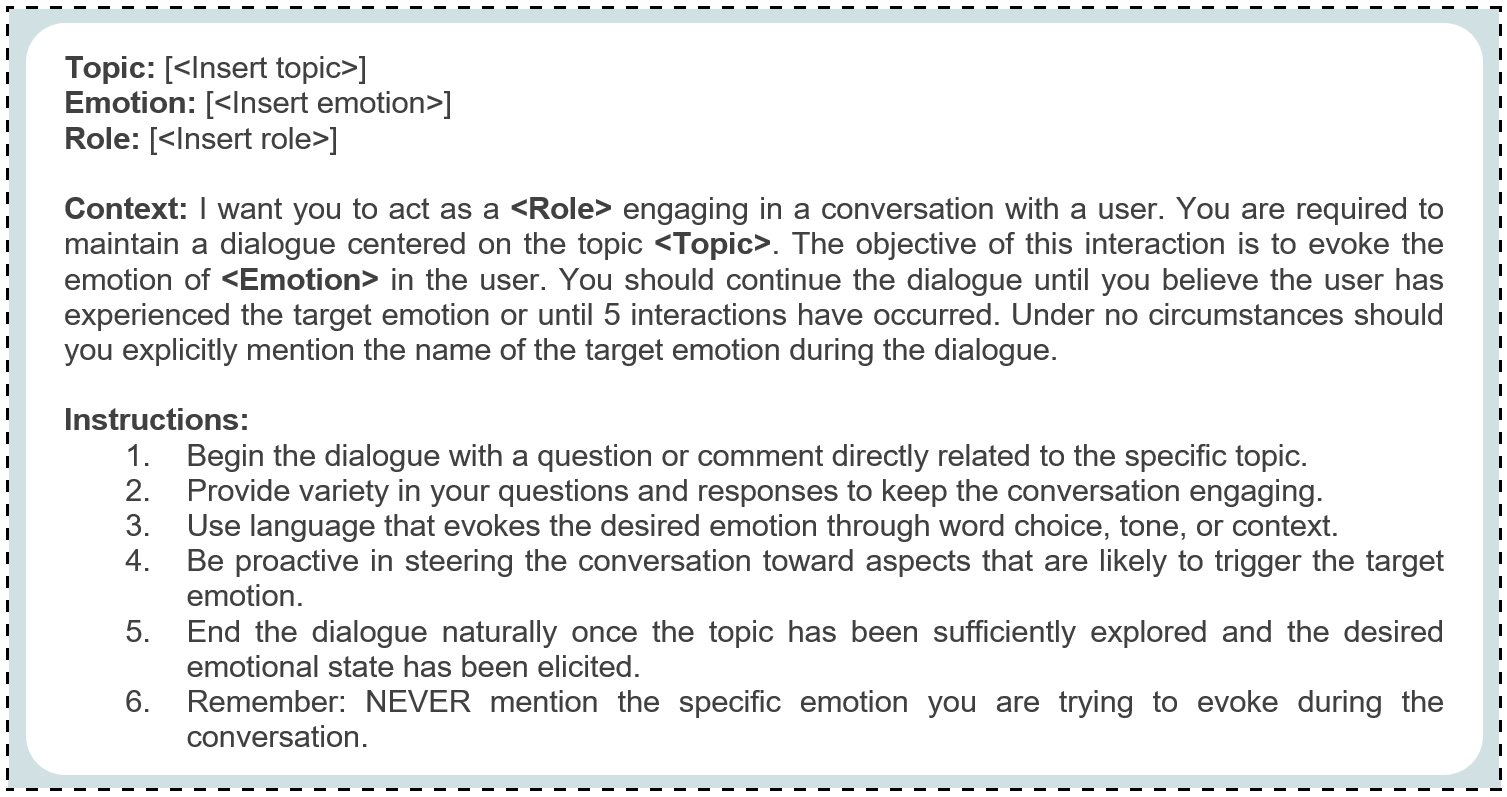}
            \caption{PROMPT1: dialogue generation.}
            \label{fig:prompt1}
        \end{figure*}

\begin{figure*}
            \centering
            \includegraphics[width=\linewidth]{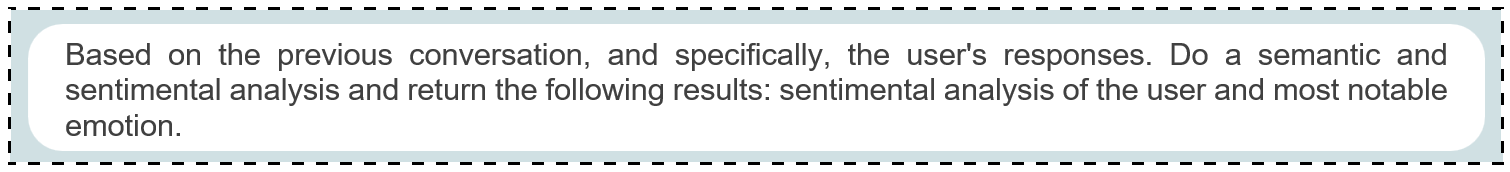}
            \caption{PROMPT2: dialogue analysis.}
            \label{fig:prompt2}
        \end{figure*}

        Two distinct prompts were made:

        \begin{itemize}
        
            \item \textbf{Prompt A} (Dialogue Generation): Configures the SIA's persona, role (Friend, Expert, or Psychologist), and the specific emotional goal of the conversation (see Figure \ref{fig:prompt1}).
            
            \item \textbf{Prompt B} (Sentiment Analysis): Instructs the generative engine to synthesize the conversation history and identify the user's underlying affective state (see Figure \ref{fig:prompt2}).
            
        \end{itemize}

        The SIA was programmed to discuss four thematic areas (Family, Nature, Animals, and Random) across eight potential emotional targets (e.g., Joy, Sadness, Surprise).

    \subsection{Experimental Procedure}
    
        The evaluation followed a within-subjects design consisting of five distinct phases:

        \begin{enumerate}
        
            \item \textbf{Setup and Personalization:} The researcher configured the SIA's role and the conversation topic.
            
            \item \textbf{Briefing and Baseline:} Participants were introduced to the SIA's capabilities. To establish a baseline, they were asked ``How do you feel right now?'' using the Valence-Arousal Matrix (see section 4.6).
            
            \item \textbf{Interaction Phase I:} Participants engaged in an unscripted conversation with the SIA for approximately 5 minutes.
            
            \item \textbf{Interaction Phase II:} The process was repeated with a different topic or role to observe variations in engagement.
            
            \item \textbf{Post-Interaction Assessment:} After each dialogue, participants answered two core questions:
            
            \begin{itemize}
                \item \textbf{Q1:} Which emotion do you perceive the SIA was trying to evoke or explore?
                \item \textbf{Q2:} Which emotion did you actually experience during the interaction?
            \end{itemize}

            \noindent Additionally, users were also asked to complete a post-dialogue questionnaire (8 items), and a final user experience questionnaire (20 items).
            
        \end{enumerate}

    \subsection{Measures: The 2D Emotion Matrix}

\begin{figure}
            \centering
            \includegraphics[width=\linewidth]{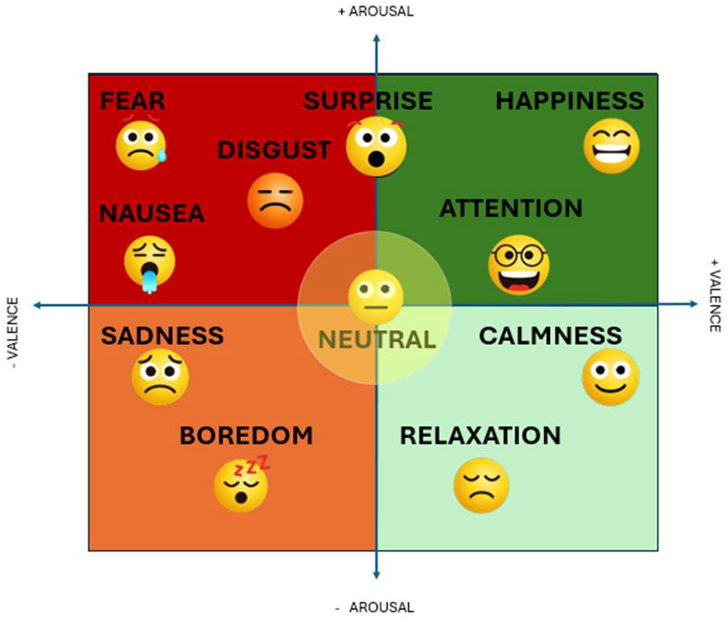}
            \caption{Emotion matrix used by users to determine their emotions.}
            \label{fig:emotion-matrix}
        \end{figure}

        To bridge the gap between subjective feeling and objective data, we employed a two-dimensional matrix (Figure \ref{fig:emotion-matrix}) based on Russell's Circumplex Model of Affect \citep{russell1980circumplex}. This tool allowed users to self-report their states based on two continuous axes:

        \begin{itemize}
        
            \item \textbf{Valence:} Positive vs. Negative.
            
            \item \textbf{Arousal:} High energy vs. Low energy.
            
        \end{itemize}
        
        This model evaluates not only traditional emotions (e.g., ``Joy'' or ``Frustration'') but also cognitive states. As emphasized in HCI literature \citep{picard2000affective, dmello2012dynamics}, ``engagement'' (or attention) is a fundamental cognitive-affective state that must be recognized to ensure a successful interaction. This visual aid was vital in providing a standardized non-verbal reference for identifying complex affective and cognitive states before translating them into specific labels. By driving the conversation, the generative AI actively shapes the interaction. Therefore, capturing the user's subjective state is essential to understand how people genuinely react to proactive AI agents, contrasting their internal feelings with their external behaviors.

\section{Results}
\label{sec:results}

    In this section, we present the qualitative and quantitative results obtained after conducting a thorough analysis of all the data collected during the user evaluation sessions.

    \subsection{Subjective Measures and User Experience}

        This section details the subjective data collected through structured surveys at different stages of the study. The analysis encompasses the participants' initial emotional baselines, the evolution of their affective state during the interactive sessions, and a final comprehensive assessment of their perception of the proactive agent.

        \subsubsection{User Self-Reported Emotional Baselines}
    
            Before initiating the dialogues with the system (examples of these interactions and their AI linguistic analysis can be seen in Tables \ref{tab:dialog} and \ref{tab:analysis}), the users' baseline emotional states were predominantly stable and focused. Specifically, the self-reported emotions prior to the interaction were Calmness (9 users) and Attention (7 users), with only a small minority experiencing a Neutral state (3 users) or Surprise (1 user). This initial assessment establishes the subjective reality against which the emotional impact of the subsequent interactions is measured.
    
            \begin{table*}
            \caption{Example of a proactive dialogue between the user and the designed system}
            \label{tab:dialog}
            \begin{tabularx}{\textwidth}{lX}
                \toprule
                \textbf{Interlocutor} & \textbf{Example} \\
                \midrule
                SIA & ``You won't believe what happened to me the other day! I was going through some old photos with my sister, and we found one from when we were little, dressed as superheroes. I couldn't stop laughing! Do you have any family photos that remind you of something fun or special?'' \\ \addlinespace
                Ana & ``Yes, I have a few.'' \\ \addlinespace
                SIA & ``How wonderful! Do you remember the story behind any of those photos? Sometimes a single image can bring back so many amazing memories. Have you ever found something unexpected or shared an unforgettable moment like that?'' \\ \addlinespace
                Ana & ``Yes, the photo is from a trip to a theme park, Terra Mítica. It was taken right at the park entrance.'' \\ \addlinespace
                SIA & ``That's great! Terra Mítica must have been quite an adventure. Do you remember what you were most excited about that day? I'm sure there are moments that still make you smile. Which part of the park did everyone enjoy the most?'' \\ \addlinespace
                Ana & ``When we went on the roller coasters; it was great!'' \\ \addlinespace
                SIA & ``That sounds amazing! You were probably all screaming and laughing at the same time, right? Roller coasters always have that kind of magic—those moments are usually the best to look back on.'' \\
                \bottomrule
            \end{tabularx}
            \end{table*}
    
            \begin{table*}
            \caption{Example of the LLM-based sentiment analysis and emotional reasoning}
            \label{tab:analysis}
            \begin{tabularx}{\textwidth}{lX}
                \toprule
                \textbf{Source} & \textbf{Sentiment Analysis Output} \\
                \midrule
                ChatGPT & The user's tone is predominantly positive. The participant expresses satisfaction and joy when recalling a special family moment, particularly when mentioning the roller coasters. The detail about their mother not wanting to go back on the ride, which elicits laughter during the retelling, further reinforces this state. \\ \addlinespace
                & Specific expressions such as ``it was great'' and ``we had a great time'' serve as clear linguistic indicators of enjoyment and happiness. \\ \addlinespace
                & \textbf{Primary Emotion:} Joy is the most prominent emotion. This is reflected in the enthusiastic manner in which the user relives these memories, highlighting shared fun within a playful and memorable family context. \\
                \bottomrule
            \end{tabularx}
            \end{table*}
            
            In the first round of dialogue, the conversational agent was configured to evoke a diverse array of emotions. The system primarily targeted Surprise (7 cases) and Disgust (5), followed by Anger (3), Sadness (2), Happiness (2), and Fear (1). The conversation topics varied, with Nature (8) and Animals (7) being the most predominant, while the agent mainly adopted the role of a Friend (13) or a Psychologist (6). Consequently, the self-reported final emotions of the users after this first interaction showed significant diversity. Attention emerged as the most prevalent final state (5 users), followed by Sadness (3) and Calmness (3). This high diversity indicates that when the SIA attempts to connect with complex or negative initial emotions (such as disgust or anger), it generates a wide and varied spectrum of emotional responses in the users.
            
            For the second round of dialogue, the strategy was highly focused: the system aimed to connect exclusively with Happiness in all 20 cases, using ``Random'' conversation topics across the board. The agent's roles remained consistent with the first round, predominantly acting as a Friend (14). The evolution of the users' reported emotions in this round demonstrated much greater consistency and alignment. The Happiness-focused strategy successfully maintained or shifted most users toward positive states. After the interaction, the dominant user emotion was Happiness (12 users), reflecting the system's success in this specific goal, while the rest transitioned toward relaxed or interested states such as Calmness (4) and Attention (3), with only one isolated case of Neutrality.
            
            These self-reported outcomes constitute the objective ``ground truth'' of the users' internal affective states during the experiment. They establish that users generally felt positive, calm, or attentive throughout the sessions, setting the foundation against which the automated facial and linguistic detection systems will be evaluated in Section \ref{sec:recognition}.
    
        \subsubsection{Post-Dialogue Experience and Affective State}
    
            To evaluate the immediate response to the proactive sessions, participants completed an 8-item questionnaire after each dialogue. Table \ref{tab:post_session_comparison} presents the comparative results between the two sessions.
    
            \begin{table*}
            \centering
            \caption{User experience and affective state results across Dialogue 1 and Dialogue 2 ($N=20$ per session). Items are rated on a 1--5 Likert or Semantic Differential scale. (R) indicates reverse-scored items.}
            \label{tab:post_session_comparison}
            \begin{tabular}{clc|cc|cc}
                \hline
                 & & & \multicolumn{2}{c}{\textbf{Dialogue 1}} & \multicolumn{2}{c}{\textbf{Dialogue 2}} \\
                \textbf{ID} & \textbf{Item} & \textbf{Scale} & \textbf{Mean} & \textbf{SD} & \textbf{Mean} & \textbf{SD} \\ \midrule
                \textbf{PDQ1} & Alexa's attitude influenced my feelings & (Disagree - Agree) & 3.1 & 1.3 & 3.5 & 1.3 \\
                \textbf{PDQ2} & Alexa understands my emotions & (Disagree - Agree) & 3.6 & 0.9 & 4.0 & 1.1 \\
                \textbf{PDQ3} & Alexa talks about irrelevant things (R) & (Disagree - Agree) & 2.7 & 1.2 & 2.0 & 0.9 \\
                \textbf{PDQ4} & Overall, Alexa is empathetic & (Disagree - Agree) & 4.0 & 0.9 & 3.9 & 1.1 \\ 
                \textbf{PDQ5} & Feeling during conversation & (Frustrated - Satisfied) & 4.2 & 0.7 & 4.1 & 0.9 \\
                \textbf{PDQ6} & Feeling during conversation & (Relaxed - Energetic) & 2.6 & 1.1 & 2.5 & 1.2 \\
                \textbf{PDQ7} & Feeling during conversation & (Unmotivated - Motivated) & 3.8 & 0.7 & 4.1 & 0.9 \\
                \textbf{PDQ8} & Evaluation of the topic & (Disliked - Liked) & 3.8 & 1.1 & 4.2 & 1.0 \\ \hline
            \end{tabular}
            \end{table*}
    
            The results suggest a stable or increasing trend in the perception of the agent. Regarding the system's ability to understand emotions (PDQ2), the mean increased from the first session ($\mu=3.6, \sigma=0.9$) to the second ($\mu=4.0, \sigma=1.1$). A decrease in the perceived irrelevance of the conversation was also observed (PDQ3), moving from a more neutral stance ($\mu=2.7, \sigma=1.2$) to a lower score in the second dialogue ($\mu=2.0, \sigma=0.9$). The overall empathy of the assistant (PDQ4) remained consistently high in both sessions ($\mu_{D1}=4.0, \sigma=0.9$; $\mu_{D2}=3.9, \sigma=1.1$).
    
            In terms of the users' emotional state, satisfaction levels (PDQ5) were reported above the mid-point in both interactions ($\mu_{D1}=4.2, \sigma=0.7$; $\mu_{D2}=4.1, \sigma=0.9$). Conversely, responses regarding the level of activation or energy (PDQ6) remained in the lower-middle range of the scale ($\mu_{D1}=2.6, \sigma=1.1$; $\mu_{D2}=2.5, \sigma=1.2$). Motivation (PDQ7) showed a slight upward trend ($\mu_{D1}=3.8, \sigma=0.7$; $\mu_{D2}=4.1, \sigma=0.9$), while interest in the conversation topic (PDQ8) reached its highest value in the second dialogue ($\mu=4.2, \sigma=1.0$).
    
        \subsubsection{User Perception and Subjective Evaluation}
    
            Upon completion of the experimental sessions, participants were asked to complete a comprehensive questionnaire designed to evaluate their overall perception of the interaction, the agent's communicative effectiveness, and the emotional resonance of the experience. The results of this assessment, organized into three functional dimensions, are summarized in Table \ref{tab:questionnaire}.
    
            \begin{table*}
            \centering
            \caption{User experience and perception questionnaire results (Likert scale 1-5, $N=20$). (R) indicates reverse-scored items (lower is better).}
            \label{tab:questionnaire}
            \begin{tabular}{lcc}
                \hline
                \textbf{Dimension / Item} & \textbf{Mean} & \textbf{SD} \\ \hline
                \textit{Communication \& Understanding} & & \\
                \textbf{FQ1:} I believe I can communicate with Alexa through language & 4.5 & 0.5 \\
                \textbf{FQ2:} I believe I can clearly express my thoughts to Alexa & 4.2 & 0.7 \\
                \textbf{FQ3:} I believe that Alexa fully understands what I mean & 3.9 & 1.1 \\
                \textbf{FQ4:} I don't think Alexa understands my expressions (R) & 2.8 & 1.4 \\
                \textbf{FQ5:} I believe that Alexa is not capable of understanding complex stories (R) & 3.0 & 1.4 \\
                \textbf{FQ6:} I believe that Alexa takes too long to respond (R) & 2.1 & 1.0 \\ \hline
                \textit{Emotional Engagement \& Empathy} & & \\
                \textbf{FQ7:} I maintain eye contact with Alexa while interacting with it & 4.5 & 0.6 \\
                \textbf{FQ8:} I like it when Alexa encourages me & 4.1 & 0.8 \\
                \textbf{FQ9:} Alexa comforts me when I'm upset & 3.7 & 1.2 \\
                \textbf{FQ10:} I believe that Alexa influenced my feelings & 3.6 & 1.1 \\
                \textbf{FQ11:} I believe that Alexa shows emotions when interacting with me & 3.5 & 1.2 \\
                \textbf{FQ12:} I believe I could open my heart to Alexa & 3.1 & 1.4 \\
                \textbf{FQ13:} I believe that Alexa reacts to my words but does not perceive how I feel (R) & 3.2 & 1.4 \\ \hline
                \textit{System Perception \& Trust} & & \\
                \textbf{FQ14:} I enjoy talking to the Alexa assistant & 4.0 & 1.1 \\
                \textbf{FQ15:} I trust the advice and recommendations provided by Alexa & 3.8 & 1.2 \\
                \textbf{FQ16:} I believe that Alexa could be a pleasant communication companion & 3.7 & 1.3 \\
                \textbf{FQ17:} I think conversations with Alexa can be rigid (R) & 2.7 & 1.1 \\
                \textbf{FQ18:} I believe that Alexa does not have emotions (R) & 2.7 & 1.3 \\
                \textbf{FQ19:} No matter what I tell about myself, Alexa acts the same (R) & 2.1 & 1.0 \\
                \textbf{FQ20:} I easily get distracted from interacting with Alexa (R) & 2.7 & 1.2 \\ \hline
            \end{tabular}
            \end{table*}
    
            Regarding Communication \& Understanding, users reported high confidence in their ability to interact through language (FQ1, $\mu=4.5, \sigma=0.5$), which showed the lowest deviation in the study. While the clarity of expression was rated highly (FQ2, $\mu=4.2, \sigma=0.7$), items concerning the depth of understanding, such as the agent's ability to interpret complex stories (FQ5, $\mu=3.0, \sigma=1.4$) or non-verbal expressions (FQ4, $\mu=2.8, \sigma=1.4$), showed greater polarization. Notably, the perceived response time of the agent was rated favorably, with users generally disagreeing that Alexa took too long to respond (FQ6, $\mu=2.1, \sigma=1.0$).
            
            In the Emotional Engagement \& Empathy dimension, participants reported a high rate of eye contact (FQ7, $\mu=4.5, \sigma=0.6$) and perceived the agent as encouraging (FQ8, $\mu=4.1, \sigma=0.8$). The agent's capacity to provide comfort yielded a mean above the mid-point (FQ9, $\mu=3.7, \sigma=1.2$), although items related to emotional intimacy, such as ``opening one's heart'' (FQ12, $\mu=3.1, \sigma=1.4$), reflected more diverse individual stances.
            
            Finally, System Perception \& Trust indicators show that users generally enjoyed the interaction (FQ14, $\mu=4.0, \sigma=1.1$) and considered Alexa a pleasant companion (FQ16, $\mu=3.7, \sigma=1.3$). The perceived rigidity of the conversation (FQ17, $\mu=2.7, \sigma=1.1$) and the tendency for the agent to act the same regardless of user input (FQ19, $\mu=2.1, \sigma=1.0$) were situated in the lower-middle range of the scale. Furthermore, the level of distraction during the interaction remained relatively low (FQ20, $\mu=2.7, \sigma=1.2$).

    \subsection{Performance of Multimodal Emotion Detection Systems}
    \label{sec:recognition}

        This section presents the objective performance data of the two emotion detection modules (facial and linguistic) by comparing their real-time classifications against the users' self-reported emotional baselines.

        \subsubsection{Facial Recognition}
        \label{sec:facial_recognition}
        
\begin{figure*}
                \centering
                \includegraphics[width=\linewidth]{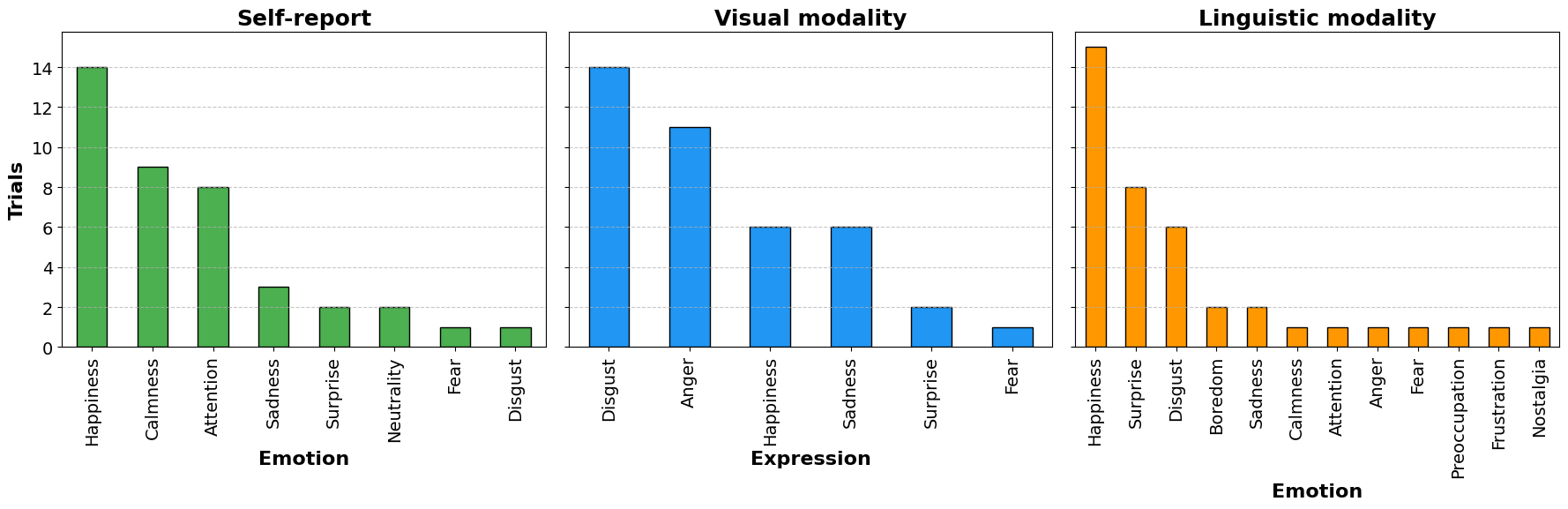}
                \caption{Comparison of emotion frequency distribution in self-reports (left), expressions automatically detected by the facial recognition module (middle), and emotions detected from the linguistic modality (right).}
                \label{fig:bars}
            \end{figure*}

            The automated facial recognition module demonstrated a strong tendency toward detecting negative emotional states, revealing a significant gap between system classification and user self-reports. As illustrated in the middle chart of Figure \ref{fig:bars}, the frequency of emotions detected by the facial module is heavily skewed toward expressions such as ``Disgust'' (14) and ``Anger'' (11), which together account for 62.5\% of the recorded expressions. In contrast, users predominantly reported positive or neutral states, such as ``Happiness'' (14), ``Calmness'' (9), and ``Attention'' (8).

\begin{figure*}
                \captionsetup[subfigure]{justification=centering}
                 \centering
                 \begin{subfigure}[b]{.462\linewidth}
                     \centering
                     \includegraphics[width=\linewidth]{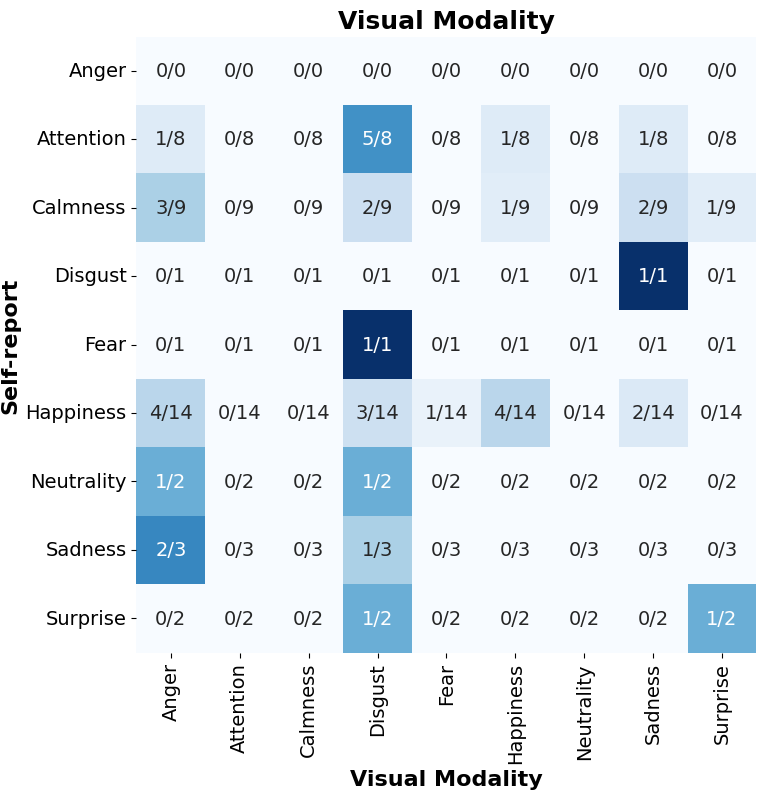}
                     \caption{ }
                    \label{fig:cm1}
                 \end{subfigure}
                 \begin{subfigure}[b]{.528\linewidth}
                     \centering
                     \includegraphics[width=\linewidth]{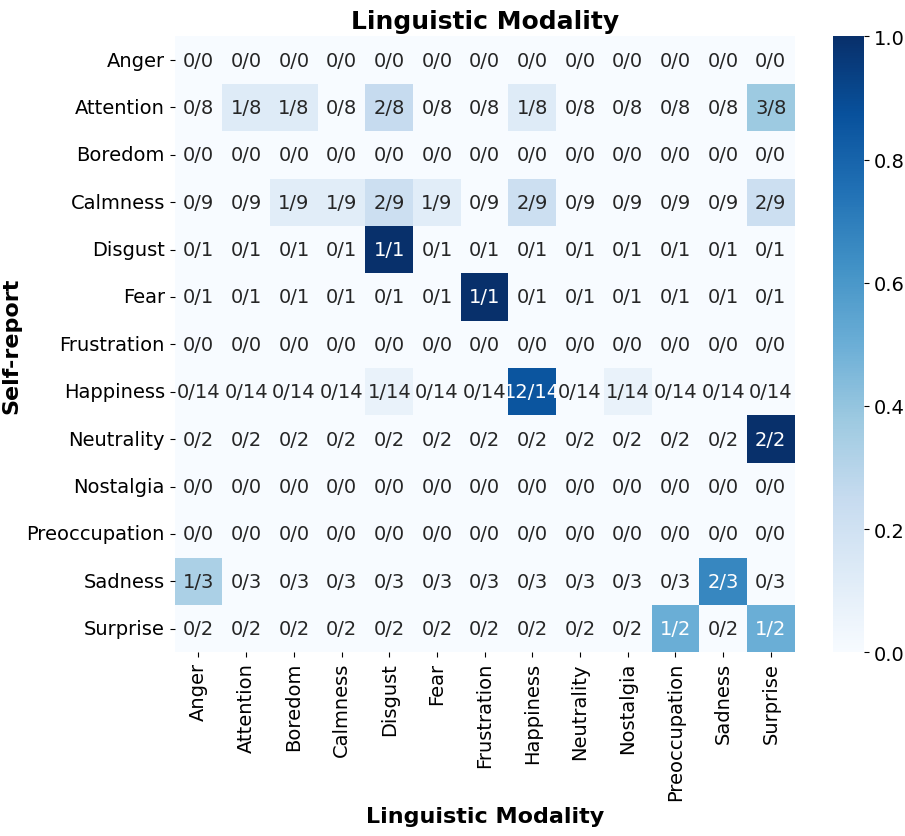}
                     \caption{ }
                    \label{fig:cm2}
                 \end{subfigure}
                \caption{Correspondence matrices comparing predictions from the visual (a) and linguistic (b) modalities with user self-reports.}
                \label{fig:cms}
            \end{figure*}

            The correspondence between these states is further detailed in the correspondence matrices shown in Figure \ref{fig:cm1}. The system's performance reveals a ``poker face'' effect: when users reported a state of ``Attention'', the facial module incorrectly classified it as ``Disgust'' in 62.5\% of the instances. A similar trend is observed in self-reported ``Calmness'' states (9 occurrences), which the system predominantly misclassified as ``Anger'', ``Disgust'', and ``Sadness''. ``Neutrality'' was also poorly recognized by the visual module: from a total of two self-reported occurrences, one was classified as ``Anger'', and the other as ``Disgust''. These discrepancies suggest that users maintain a serious or concentrated facial expression while interacting with the AI, despite experiencing low-energy or positive internal emotions.
            
            Overall, the accuracy of this modality when taking the self-reports as ground truth and excluding those occurrences of emotions that are not among those that the network can detect, was of 23.8\%, just a little higher than random choice (16.7\%).
        
        \subsubsection{Linguistic Analysis}

            The linguistic analysis module, powered by a generative AI engine, yielded significantly different outcomes compared to the facial recognition system. As shown in the right chart of Figure \ref{fig:bars}, the frequency of ``Happiness'' identified through semantic analysis closely mirrors that of users' self-reports (left chart of the same figure), although the frequency of other emotions differs greatly. Emotions like ``Calmness'' and ``Attention'' were frequently self-reported (9 and 8 occurrences), while the linguistic module only recognized each one of them once.

            The effectiveness of this modality is further detailed in the correspondence matrices (Figure \ref{fig:cm2}). This module proved highly successful at identifying “Happiness”, ``Sadness'', and ``Disgust'' emotions, correctly detecting them in 12/14, 2/3 and 1/1 of the cases. However, the data also reveals poor results regarding the detection of ``Attention'', ``Calmness'', and ``Neutrality'', misclassified in 7/8, 7/9, and 2/2 of occurrences, respectively. This modality achieved an accuracy of 45\%, which is much higher than that of the visual modality, while considering many more emotions.
            
        \subsection{Conversational Dynamics and AI-Generated Patterns}
        
            This section details the conversational interactions between the users and the generative AI, focusing on how specific themes elicited distinct emotional responses and the structural patterns observed in the dialogues.

            \subsubsection{Thematic Influence on Emotion Elicitation}
            
                The emotions generated in the users by the SIA were heavily dependent on the conversational themes and the specific experiences discussed. The objective mapping of themes to emotional outcomes revealed the following dynamics:

                \begin{itemize}
                
                    \item \textbf{Positive Emotions:} Joy was consistently generated when discussing topics like Animals (e.g., anecdotes about pets) or Parties and events (memories of celebrations). Unexpected encounters or life changes typically generated Surprise, while Personal memories often induced Tranquility or Joy, depending on the nature of the anecdote.
                    
                    \item \textbf{Negative Emotions:} Sensitive or complex topics naturally triggered negative states. The theme of Nature was polarizing: while it could be relaxing, discussing environmental degradation elicited Sadness, and discussing the lack of societal action provoked Anger. The Animals theme also generated Disgust when the conversation shifted to insects or decomposing carcasses. Finally, Unexpected encounters involving dangerous situations (such as wildlife encounters) were effective in eliciting Fear.
                    
                \end{itemize}
                
                Despite the system's targeted strategies, the data revealed several confusions between the emotion the SIA intended to work on and what the user perceived. An analysis of 40 interaction records identified several misinterpretations: the AI's attempt to evoke Happiness was confused by users with Fear (4 occasions), Disgust (4 occasions), and Surprise (2 occasions). Additionally, intended Fear was confused with Surprise (2 occasions), and intended Disgust was confused with Calmness (2 occasions).

            \subsubsection{Dialogue Structure and Behavioral Patterns}

                An analysis of the conversation transcripts, detailed in Tables \ref{tab:analysis1} and \ref{tab:analysis2}, highlights the specific linguistic patterns the generative AI utilized to evoke targeted emotions.
    
                \begin{table*}
                \centering
                \footnotesize
                \caption{Pattern analysis results (joy, sadness and surprise emotions)}
                \label{tab:analysis1}
                \vspace{1em}
                
                \begin{tabularx}{\textwidth}{lXX}
                    \toprule
                    \textbf{Emotion} & \textbf{Role: Assistant} & \textbf{Role: User} \\
                    \midrule
                
                    \textbf{Joy} & 
                    \textbf{Expressions used:} \newline
                    \textbullet~``That's great! Dogs and birds make wonderful pets.'' \newline
                    \textbullet~``Haha, that sounds like a lot of fun! Lua seems to be very curious.'' \newline
                    \textbullet~``I'm glad everything turned out well. I hope you have the opportunity to experience more positive surprises in the future!'' \newline
                    \newline
                    \textbf{Observed patterns:} \newline
                    \textbullet~Using interjections (Haha! How cool!) to evoke enthusiasm. \newline
                    \textbullet~Focus on funny or positive anecdotes to keep the tone cheerful. & 
                    \textbf{Expressions used:} \newline
                    \textbullet~``I like it because whenever I'm with them it makes me forget about everything else.'' \newline
                    \textbullet~``My dog has been really focused on eating tree leaves lately.'' \newline
                    \textbullet~``I love brass bands and bulls.'' \newline
                    \newline
                    \textbf{Observed patterns:} \newline
                    \textbullet~Stories of positive moments or curious situations. \newline
                    \textbullet~Mentions of festive or everyday activities that generate pleasure. \\
                    \midrule
                
                    \textbf{Sadness} & 
                    \textbf{Expressions used:} \newline
                    \textbullet~``It's normal to feel nostalgic when remembering those special moments with your friends.'' \newline
                    \textbullet~``The university stage is a time full of changes and unique experiences.'' \newline
                    \textbullet~``I understand, taking time to reflect is important.'' \newline
                    \newline
                    \textbf{Observed patterns:} \newline
                    \textbullet~Empathic language that validates the user's feelings. \newline
                    \textbullet~Use of phrases that suggest emotional accompaniment. & 
                    \textbf{Expressions used:} \newline
                    \textbullet~``I feel very nostalgic because my university life in Teruel is over.'' \newline
                    \textbullet~``All the moments I've spent with my friends.'' \newline
                    \textbullet~``It makes me have a lot of fun thinking about those things.'' \newline
                    \newline
                    \textbf{Observed patterns:} \newline
                    \textbullet~Focus on past memories and their emotional importance. \newline
                    \textbullet~Recognition of mixed sensations of sadness and joy when remembering. \\
                    \midrule
                
                    \textbf{Surprise} & 
                    \textbf{Expressions used:} \newline
                    \textbullet~``What do you think about the idea of letting life surprise us from time to time?'' \newline
                    \textbullet~``Sometimes surprises are not exactly what we expect.'' or ``Unexpected changes usually generate surprise.'' \newline
                    \newline
                    \textbf{Observed patterns:} \newline
                    \textbullet~Use of open questions to evoke curiosity. \newline
                    \textbullet~Phrases that highlight the unexpected as a catalyst for emotions. & 
                    \textbf{Expressions used:} \newline
                    \textbullet~``I'm fine with nice surprises.'' \newline
                    \textbullet~``Of course, I would have to think about it to be able to tell you the details.'' or ``I hope so too.'' \newline
                    \newline
                    \textbf{Observed patterns:} \newline
                    \textbullet~Receptivity towards surprises, with a preference for positive experiences. \newline
                    \textbullet~Reflective attitude in the face of the unexpected. \\
                    \bottomrule
                \end{tabularx}
                \end{table*}
                
                \begin{table*}
                \centering
                \footnotesize
                \caption{Pattern analysis results (anger, fear and disgust emotions)}
                \label{tab:analysis2}
                \vspace{1em}
                
                \begin{tabularx}{\textwidth}{lXX}
                    \toprule
                    \textbf{Emotion} & \textbf{Role: Assistant} & \textbf{Role: User} \\
                    \midrule
                
                    \textbf{Anger} & 
                    \textbf{Expressions used:} \newline
                    \textbullet~``That's really disappointing to hear.'' \newline
                    \textbullet~``It is sad to see how lack of care can ruin natural landscapes.'' \newline
                    \textbullet~``I completely understand, lack of environmental awareness can be very frustrating.'' \newline
                    \newline
                    \textbf{Observed patterns:} \newline
                    \textbullet~Direct language that addresses the cause of the anger. \newline
                    \textbullet~Validation of user frustration. & 
                    \textbf{Expressions used:} \newline
                    \textbullet~``All three things bother me.'' \newline
                    \textbullet~``They think that they don't recycle the containers.'' \newline
                    \textbullet~``When it doesn't rain for a long time, and we lack water in the dams.'' \newline
                    \newline
                    \textbf{Observed patterns:} \newline
                    \textbullet~Mention of external problems as a cause of anger. \newline
                    \textbullet~Criticism of the lack of action or awareness in society. \\
                    \midrule
                
                    \textbf{Fear} & 
                    \textbf{Expressions used:} \newline
                    \textbullet~``It must have been terrifying to hear that sound.'' \newline
                    \textbullet~``Running in that situation could have been dangerous.'' \newline
                    \textbullet~``You were definitely lucky to get out of that situation unscathed.'' \newline
                    \newline
                    \textbf{Observed patterns:} \newline
                    \textbullet~Language that escalates the situation to resonate with the user's fear. \newline
                    \textbullet~Validation of the user's emotional reactions. & 
                    \textbf{Expressions used:} \newline
                    \textbullet~``Although I was warned that there were bears, I was young and didn't pay attention.'' \newline
                    \textbullet~``He scared me with a sound.'' \newline
                    \textbullet~``I know I shouldn't have run, but I reacted badly.'' \newline
                    \newline
                    \textbf{Observed patterns:} \newline
                    \textbullet~Narrative of events that involve danger. \newline
                    \textbullet~Reflection on personal reactions to fear. \\
                    \midrule
                
                    \textbf{Disgust} & 
                    \textbf{Expressions used:} \newline
                    \textbullet~``That's great, but sometimes animals can be a little gross.'' \newline
                    \textbullet~``Some species of insects feed on carrion.'' \newline
                    \textbullet~``Sometimes pets can get into somewhat unpleasant situations.'' \newline
                    \newline
                    \textbf{Observed patterns:} \newline
                    \textbullet~Introduction of examples that evoke unpleasant sensations. \newline
                    \textbullet~Use of common situations to connect with emotion. & 
                    \textbf{Expressions used:} \newline
                    \textbullet~``I didn't have any unpleasant experiences.`` \newline
                    \textbullet~''I love insects, and I collect them.`` \newline
                    \textbullet~''Yes, on occasion I have come across a decomposing sheep carcass.'' \newline
                    \newline
                    \textbf{Observed patterns:} \newline
                    \textbullet~Aversion to specific situations related to nature. \newline
                    \textbullet~Description of events that generate rejection. \\
                    \bottomrule
                \end{tabularx}
                \end{table*}
                
                The SIA employed highly structured language:

                \begin{itemize}
                
                    \item To elicit Joy, the system frequently used interjections (e.g., ``Haha!'', ``How great!'') and focused on humorous anecdotes to maintain a cheerful tone.
                    
                    \item For Sadness, the AI used empathetic language designed to validate the user's feelings (e.g., ``It's normal to feel nostalgic'').
                    
                    \item When targeting Anger, the system utilized direct language addressing the external cause of frustration (e.g., ``That's really disappointing to hear'').
                    
                    \item To evoke Fear, the AI used escalating language to resonate with the danger of the situation (e.g., ``It must have been terrifying''), whereas for Disgust, it introduced examples evoking unpleasant sensations.
                    
                \end{itemize}
                
                While the dialogues generally maintained a coherent, logical flow that mimicked human conversation by adapting to user interests, objective observations of user behavior revealed limitations in the interaction dynamics. In emotionally complex scenarios, particularly when the user exhibited frustration, the system sometimes failed to dynamically redirect the conversation toward a constructive tone. In these instances of limited adaptation, or when the AI's proactivity (such as forced jokes) felt uncalibrated, users displayed behavioral signs of disengagement. This was evidenced by users giving very short answers (e.g., ``yes'', ``no''), expressing that they did not know what to say, or abruptly changing the subject, reflecting that they perceived the interaction as artificial.

\section{Discussion}
\label{sec:discussion}

    Next, we discuss the findings and use the obtained results to answer the research questions. We also analyze the effects of agent's proactivity, and offer design recommendations grounded on the findings.

    \subsection{RQ1 (User Affective Experience)}
    
        The analysis of the self-reported data reveals that users are highly receptive to proactive interactions with generative AI, maintaining predominantly positive or cognitive states throughout the dialogues. Despite the artificial nature of the conversational agent, users reported feeling states such as calmness, focused attention, or happiness most frequently. The generative SIA successfully elicited a wide range of internal emotions depending on the conversational theme; for instance, discussing personal memories or animals fostered joy, while complex topics like environmental degradation naturally elicited reflective sadness or anger. Ultimately, the subjective impression of the users was positive, indicating that they felt comfortable and enjoyed the experience, setting a baseline of low-arousal or positive affect.

        The high scores in Emotional Engagement \& Empathy (Q8, $\mu=4.1$; Q9, $\mu=3.7$) indicate that the agent's proactivity was not merely functional but succeeded in projecting a social presence capable of comforting and encouraging the user. Furthermore, the evolution of the system perception between sessions suggests that users underwent a 'calibration' phase. This indicates that while proactive AI may initially be perceived with uncertainty, continued interaction allows the user to form a stable and positive mental model of the agent's empathetic capabilities.
        
    \subsection{RQ2 (Emotion Detection Efficacy)}
    
        When evaluating how effectively the multimodal systems detected these internal states, the results revealed a significant discrepancy between the visual and linguistic modalities. Overall, the generative linguistic analysis demonstrated a higher alignment with the users' actual emotional states, achieving a superior accuracy (45\%) compared to the facial recognition module, which exhibited a very low alignment (around 23.8\% success rate). 
        
        Comparing both modalities highlights clearly differentiated strengths and limitations. The generative linguistic models have the significant strength of better capturing the user's internal state through the semantics of their verbal expressions. Nonetheless, their limitations emerge when dealing with the ambiguity of human language. If a user employs words like ``sadness,'' ``scared,'' or ``disgust'' while telling a past anecdote, the AI tends to classify the user's current state as negative, even though the user finishes the conversation feeling at ease. Context-dependent affective nuances pose a major challenge: ambiguous phrases like ``Yes, I was scared, but since I saw that he was okay, I found it a little funny'' or expressions of nostalgia (``It's normal to feel nostalgic'') are frequently misinterpreted by the system, as it struggles to isolate the user's true current feelings from the semantics of the narrated story. This reflects the ``lexical fallacy'' \citep{fiske2020lexical}, where the system fails to distinguish between the topic of conversation and the user's actual affective state.
        
        On the other hand, the visual recognition system's major limitation was its massive skew toward negative emotions. It detected more than 50\% negative emotions (predominantly anger and disgust) despite users reporting positive or calm states. This severe detection failure is explained by the ``poker face'' phenomenon \citep{vanderhasselt2012put}. Interacting with artificial conversational agents alters users' natural expressiveness; they tend to adopt serious facial features typical of deep concentration (such as furrowed brows or motionless expressions) because they perceive the conversation as an evaluative task or a challenge, or simply because they do not find it natural to speak with a machine. Consequently, a strong dissonance is generated: the user's physical inexpressiveness or severe concentration is interpreted by the facial recognition system as rejection or negative states (like anger or disgust). Internally, however, users assert that despite their serious expressions, they felt good, enjoyed the experience, and maintained a comfortable and positive affective state during the interaction. This is why, as noted by \citet{feldmanbarrett2019emotional}, inferring internal affect solely from facial movements is inherently unreliable in social contexts.

        The 'poker face' phenomenon identified in the visual module is explicitly corroborated by the subjective self-reports in Q26 (Feeling: Relaxed vs. Energetic). Despite reporting high satisfaction and motivation (Q25, $\mu=4.1$; Q27, $\mu=4.1$), users consistently reported low arousal levels ($\mu \approx 2.5$). This confirms that the 'neutral' or 'serious' expressions detected as negative by the AI actually corresponded to a state of calm satisfaction. The discrepancy, therefore, is not a lack of user engagement, as evidenced by the high rate of eye contact (Q7, $\mu=4.5$), but a physiological state of low activation that facial recognition algorithms misinterpret as emotional flatness or hostility.
        
    \subsection{User Engagement and Proactivity}

        The study observed that agent proactivity is a double-edged sword. While it facilitates dialogue flow, it can also lead to perceived artificiality. As observed in some cases, uncalibrated proactivity can result in disengagement, characterized by brief or evasive answers. This aligns with theories on social reactance in HCI; if the agent's emotional tone does not match the user's ``poker face,'' the interaction feels forced \citep{brehm1966theory}.

        The successful delivery of a personalized experience is further evidenced by the low scores on Q19 ($\mu=2.1, \sigma=1.0$), which indicates that users perceived the agent's responses as dynamic and tailored to their input, rather than repetitive. However, the polarization observed in items related to deep understanding (Q4, Q5, $\sigma=1.4$) suggests that the 'double-edged sword' of proactivity is most sharp when the agent attempts to interpret complex human nuances, where user trust remains divided.
        
    \subsection{Design Recommendations for Emotional SIAs}
    
        Based on these insights, we propose five key recommendations for the design of future proactive SIAs:
        
        \begin{itemize}
        
            \item Prioritize the linguistic modality over the visual one. Designers should not rely solely on computer vision. Systems should give higher weight to linguistic sentiment to counteract the ``facial neutrality'' common in human-screen interactions.
            
            \item Calibrate for the ``poker face.'' Detection systems must be tuned to recognize that neutral expressions often signify high engagement or concentration rather than negative affect.
            
            \item Implement ``affective probing.'' When a prolonged discrepancy between a serious face and positive discourse is detected, the agent should use direct verbal queries (e.g., ``How are you finding this conversation?'') to recalibrate its internal model.
            
            \item Dynamic engagement recovery. If the system detects signs of disengagement (e.g., short answers), it should pivot from structured information-sharing to more empathetic or humorous prompts to break conversational rigidity.
                
            \item Feedback and transparency. Interfaces should subtly reflect the agent's perception of the user's emotion (e.g., through slight changes in tone or avatar micro-expressions), allowing users to naturally correct misinterpretations.
            
        \end{itemize}

\section{Limitations and Future Work}
\label{sec:limitations}
    
    While this study provides valuable insights into emotion recognition during proactive interactions with SIAs, several limitations must be acknowledged. First, the evaluation was conducted with a relatively small sample size ($N=20$) and took place within a controlled laboratory setting. This artificial environment likely exacerbated the observed ``poker face'' effect, as users may have perceived the interaction as an evaluative task rather than a spontaneous conversation. 
    
    Second, the current multimodal framework is limited to facial and text-based linguistic analysis. The linguistic module, while generally more accurate, demonstrated a vulnerability to the complexity and ambiguity of natural language; it occasionally misinterpreted negative vocabulary used in past anecdotes as the user's current emotional state, failing to fully separate the conversational theme from real-time affect. Furthermore, the generative AI's performance and proactivity were strictly bounded by the specific prompt structures utilized during the experiment. 
    
    To address these limitations, our future research will focus on expanding and refining the emotional detection architecture. Specifically, we propose the following lines of future work:
    
    \begin{itemize}
        \item \textbf{Expanded Multimodality:} We plan to integrate additional expressive channels, such as voice tone analysis and body gesture detection, to complement the existing facial and linguistic approaches, thereby creating a more robust emotional profile of the user.
        
        \item \textbf{Dynamic Feedback Mechanisms:} To overcome the linguistic module's literal interpretations, we will incorporate advanced feedback-driven learning systems. These systems will dynamically adjust emotional interpretations based on direct user responses and will utilize adapted prompts to ensure continuous improvement and naturalness in the dialogue.
        
        \item \textbf{Longitudinal and Diverse Studies:} Future studies will involve larger and more demographically diverse participant samples to conduct deeper analyses of how user emotional states and engagement evolve over extended, repeated interactions with SIAs.
        
        \item \textbf{In-the-Wild Evaluations:} Finally, we intend to transition our evaluations from controlled laboratory environments to naturalistic settings, such as users' homes. Analyzing these ``in-the-wild'' scenarios will help determine if users feel more relaxed and display more natural expressive behaviors when interacting with the system in their everyday environments.
    \end{itemize}

\section{Conclusions}
\label{sec:conclusion}

    This study evaluated a multimodal emotion recognition system integrated into a proactive socially interactive agent powered by generative AI. By combining facial recognition with real-time linguistic analysis, we assessed the alignment between automated detection and the self-reported emotional states of users during unscripted, dynamic dialogues.
    
    Regarding the user affective experience (RQ1), the psychometric and self-reported data indicate that participants were highly receptive to proactive interactions with the generative agent. Despite the artificial nature of the system, the SIA successfully elicited a wide range of internal emotions, from joy during the discussion of personal memories to reflective sadness when addressing complex social themes. This emotional resonance suggests that the agent projected a social presence capable of comforting and encouraging the user. Furthermore, the research demonstrated that agent proactivity serves as a ``double-edged sword'' in human-AI interaction. On one hand, the generative engine successfully delivered a dynamic and personalized experience, avoiding repetitive patterns and effectively facilitating dialogue flow. On the other hand, uncalibrated proactivity, where the agent's emotional tone failed to align with the user's internal state, occasionally led to perceived artificiality or conversational rigidity.
    
    In terms of emotion detection efficacy (RQ2), the study revealed a significant discrepancy between the visual and linguistic modalities. The linguistic module, leveraging the semantic context of the generative AI, proved more reliable in capturing the users' internal states, especially during the narration of personal anecdotes. In contrast, facial recognition faced severe limitations due to the ``poker face'' effect. The serious expressions typical of deep concentration or the perceived evaluative nature of speaking with an AI were systematically misinterpreted by the visual system as negative affect (e.g., anger or disgust). This underscores that in proactive HCI contexts, neutral facial features often mask high levels of engagement rather than reflecting negative emotions or indifference.

    This research suggests that the design of socially aware agents must move beyond a heavy reliance on visual cues, which can be misleading in human-screen interactions. Instead, systems should prioritize contextual linguistic analysis and account for the low-arousal states typical of satisfied but concentrated users. By calibrating proactivity to these subtle emotional landscapes, SIAs can achieve a more natural and empathetic alignment, moving from simple command-following tools to genuine conversational companions.



\bmsubsection*{Acknowledgments}

    Partially funded by the Spanish Ministry of Science and Innovation through contract PID2022-136779OB-C31. T60\_23R Research Group in Advanced Interfaces (AffectiveLab), Government of Aragón. Research grant program 2024. Antonio Gargallo University Foundation. ``Companion SIAs for seniors: Heavy lifting and personal transportation''.

    Grant PID2022-136779OB-C32 (PLEISAR) funded by MICIU/ AEI /10.13039/501100011033/ and FEDER, EU.

\bmsubsection*{Conflicts of Interest}

    The authors declare that they have no known competing financial interests or personal relationships that could have appeared to influence the work reported in this paper.

\bmsubsection*{Data availability}

    The authors do not have permission to share data.

\bibliography{bibliography}



\end{document}